\documentstyle[12pt]{article}

\def \e {equation}
\def \n {\noindent}
\begin{document}

\centerline {\bf NON-LINEAR APPROXIMATIONS TO GRAVITATIONAL INSTABILITY:}

\centerline {\bf A COMPARISON IN THE QUASI-LINEAR REGIME}

\bigskip
\bigskip
\centerline{Dipak Munshi$^1$, Varun Sahni$^1$}
\bigskip
\centerline{and}
\bigskip
\centerline{Alexei A. Starobinsky$^{2,3}$}
\medskip
\centerline{$^1$ Inter-University Centre for Astronomy and Astrophysics}
\centerline{Post Bag 4, Ganeshkhind, Pune 411 007, India}
\vskip .2cm
\centerline{$^2$ Yukawa Institute for Theoretical Physics, Kyoto University,
Uji 611, Japan}
\centerline{$^3$ Landau Institute for Theoretical Physics, Russian Academy
of Sciences,}
\centerline{ Moscow 117334, Russia (permanent address)}
\bigskip
\eject
\topskip 1.0cm
\centerline{\bf Abstract}
\vskip .5cm
We compare different nonlinear approximations to gravitational clustering
in the weakly nonlinear regime, using as a comparative statistic the
evolution of non-Gaussianity  which can be characterised by a set of numbers
$S_p$ describing connected moments of the density field
at the lowest order in $\langle\delta^2\rangle$:
$\langle\delta^n\rangle_c \simeq S_n\langle\delta^2\rangle^{n-1}$.
Generalizing earlier work by Bernardeau (1992) we develop an ansatz to
evaluate all $S_p$ in a given approximation by means of a generating
function which can be shown to satisfy the equations of motion of a
homogeneous spherical density enhancement in that approximation. On the
basis of the values of we show that approximations formulated in
Lagrangian space (such as the
Zeldovich approximation and its extensions) are considerably more accurate
than those formulated in Eulerian space such as the Frozen Flow and Linear
Potential approximations. In particular we find that the $n$th order Lagrangian
perturbation approximation correctly reproduces the first $n+1$ parameters
$S_n$. We also evaluate the density probability distribution function
for the different approximations in the quasi-linear regime
and compare our results with an exact analytic treatment in the case of
the Zeldovich approximation.

\bigskip

\n
{\it Subject Headings: cosmology: theory -- large scale structure of Universe}
\eject
\bigskip
\centerline{\bf 1. Introduction}
\bigskip
\n
Gravitational instability in the Universe can be characterised by two
distinct epochs: during the first, density fluctuations
$\delta \equiv (\rho - \rho_0)/\rho_0$ evolve self-similarly
according to the tenets of linear theory
($\delta^{(1)}({\bf x},t) \propto D_+(t)\Delta({\bf x})$)
with the result that a density distribution
that was initially Gaussian remains Gaussian at subsequent epochs as long
as $|\delta| << 1$. The linear epoch clearly cannot continue indefinitely
since a stage will arise when $\delta$, although small, becomes
comparable to unity, so that weakly nonlinear effects can no longer be ignored.
This is the quasi-linear regime; as long as $|\delta| < 1$
the evolution of the density field can be adequately described
by means of a perturbative expansion:
$\delta (\vec x, t) = \sum_{n=1}^{\infty}\delta^{(n)}({\bf x},t)$ where
$\delta^{(n)}({\bf x}, t) = D_+^{n}(t)\Delta^{(n)}({\bf x})$
if the Universe is spatially flat and matter dominated (Peebles 1980,
Fry 1984).
This epoch witnesses the growth of skewness and
kurtosis and other higher moments of the one-point probability
distribution function (PDF) of density perturbations
characterising a non-Gaussian development
of an initially Gaussian field $\delta^{(1)}$.
During still later stages, $\delta$ becomes larger than unity,
with the result that perturbative approximations break down
and a fully nonlinear treatment of the problem is required.
This later epoch is characterised by the formation of pancakes
($\delta \rightarrow \infty$ along two dimensional sheets)
and the gradual development
of cellular structure (Shandarin $\&$ Zeldovich 1989).

Although no exact treatment is available
which  encompasses the entire
nonlinear epoch, some approximations have been suggested which
attempt to mimick certain features of nonlinear gravitational
instability. Our aim in this paper is to investigate the relative
accuracy of five such approximations in the weakly nonlinear regime
($|\delta |< 1$) where they may be analytically compared to the exact
solution in the form of a perturbative expansion in powers of $\delta$.

Our treatment overlaps with (and considerably extends) the recent work of
Munshi $\&$ Starobinsky (1993) (hereafter MS) and Bernardeau et al. (1993),
in which 3 approximations: the Zeldovich approximation (Zeldovich 1970),
hereafter ZA, the frozen flow approximation
(Matarrese et al. 1992), hereafter FF, and the linear potential approximation
(Brainerd et al. 1993; Bagla \& Padmanabhan 1994), hereafter LP, were
compared to the exact solution in second and third order in
perturbation theory. In this paper, we provide an ansatz whereby
nonlinear approximations can be compared with the exact perturbative solution
to any order in perturbation theory. We apply this ansatz
to approximations formulated in Eulerian space such as FF and LP, as well as
to approximations based on Lagrangian perturbation theory (of which ZA happens
to be the lowest order solution). It appears that Lagrangian perturbative
methods are, as a rule, much more accurate than either FF or LP to
{\bf any given order} in perturbation theory.

\bigskip
\centerline{\bf 2. Nonlinear approximations}
\bigskip
\n
Gravitational instability in a spatially flat matter dominated FRW Universe
before caustic formation can, in the Newtonian approximation,
be described by the following system of
equations (see, e.g., Zeldovich \& Novikov 1983, Peebles 1980):
$$\triangle_{\bf x} \Phi =4\pi Ga^2\rho_0\delta;~~~~ \eqno(1a)$$
$$\dot \delta + {1\over a} \nabla_{\bf x}.~((1+\delta){\bf u})=0; \eqno(1b)$$
$$(a{\bf u})^{\cdot}+({\bf u} \nabla_{\bf x}){\bf u} =
-\nabla_{\bf x} \Phi \eqno(1c)$$
where ${\bf u} = a\dot{\bf x}$, ${\bf x}$ being the comoving coordinate
($a(t)\propto t^{2/3}$ and $\rho_0=1/6\pi Gt^2$). If
we assume that ${\bf u}$ is irrotational then we can define a
velocity potential such that ${\bf u}({\bf x},t)= \nabla_{\bf x}v({\bf x},t)
\equiv \nabla U({\bf x},t)$ where we have introduced the notation
$\nabla \equiv {1\over aH} \nabla_{\bf x}$, $H=\dot a/a=2/3t$
(the potential $U$ is related
to the potential $V$ used in MS by the relation $U=-HV$). Moreover,
$\triangle U= {1\over aH}~ div_{\bf x} {\bf u} =\theta$ which is the
dimensionless velocity divergence used in MS.

In this notation, we can rewrite the equations as follows:
$$ a{\partial \over \partial a}\delta ({\bf x},a)+ (1 + \delta({\bf x},a))
\theta ({\bf x},a)+ \nabla \delta({\bf x},a)\nabla U({\bf x},a)
= 0; \eqno(2a)$$
$$ a{\partial \over \partial a}\theta({\bf x},a) + {1\over 2 }\theta
({\bf x},a) + \nabla U({\bf x},a) \nabla \theta ({\bf x},a)) +
U_{ik}U_{ik} + \triangle \Phi
({\bf x},a) =0; \eqno(2b)$$
$$\triangle \Phi = {3\over 2}\delta ({\bf x}, a) \eqno(2c)$$
where $U_{ik} = ({1\over {aH}})^2{{\partial}^2}_{ik}U$
and summation over repeated indexes is assumed. The initial
condition for the system (2) at $t \to 0$ is given by the linear
approximation corresponding to a growing adiabatic mode:
$$ \Phi = \phi_0 ({\bf x});\,~ U=-{2\over 3}\phi_0 ({\bf x});\, ~\delta =
{2\over 3}\triangle \phi_0 ({\bf x});\, ~\theta = - \delta \, .\eqno(3)$$

Several of the nonlinear approximations which we will consider such as
the Zeldovich approximation, the Frozen Flow
approximation and the Linear Potential approximation arise
as a result of the extrapolation of one of the linearised relations in (3)
into the nonlinear regime. Thus in FF, the value of the
velocity potential is kept fixed to its linearised value
$U = -{2\over 3}\phi_0$, so that ${\bf u} = - {2\over 3}\nabla \phi_0$.
The movement of a particle in FF is such that the particle
upgrades its velocity with each time step to that prescribed by the local
value of the linear velocity field. This results in a laminar flow for the
velocity field since different particle trajectories can
never intersect in principle.

The linear potential approximation on the other hand, is based upon
the assumption that the gravitational potential does not evolve with
time so that
$ \Phi = \phi_0({\bf x})$. As a result particles effectively move along the
lines of force of the primordial potential $\phi_0$.
Similarly the Zeldovich approximation is based on extending
$ U = -{2\over 3}\Phi$ into the nonlinear regime.
In all three cases the constraint equations
$ U = -{2\over 3}\Phi$~(ZA),
$ U = -{2\over 3}\phi_0$~(FF),
$ \Phi = \phi_0 $~(LP)
replace the Poisson equation (2c) which is not satisfied in these
approximations beyond the linear regime.

In nonlinear approximations formulated in Lagrangian space,
the main object of study is the particle trajectory. In these approximations
the initial comoving (Lagrangian) coordinate $\bf q$ and the Eulerian
coordinate of a particle
${\bf x(t)}$ are related by a displacement field ${\bf \Psi}$:
$${\bf x} = {\bf q} + {\bf \Psi}(t, \bf q).\eqno(4)$$
If we introduce the matrix
$$M_{ik}(t, {\bf q})={\partial x_i\over \partial q_k}=\delta_{ik}+
{\partial \Psi_i\over \partial q_k},\eqno(5)$$
then $M_{ik}$ satisfies the equations
$${\partial \over \partial t}\left(a^2{\partial M_{ik}\over \partial t}
\right)\cdot M^{-1}_{ki}+{2a^2\over 3t^2}(J^{-1}-1)=0,\eqno(6a)$$
$$\epsilon_{ikl}\dot M_{km}M^{-1}_{ml}=0,\eqno(6b)$$
where $J=|det \lbrack M_{ik}\rbrack |$ is the Jacobian of the
transformation (4), $\epsilon_{ikl}$ is the unit totally
antisymmetric tensor, Eq. (6b) being the condition of potential motion
in Eulerian space (see, e.g., Zeldovich \& Novikov 1983). Density and
velocity fields are determined from the relations
$${\rho\over \rho_0}\equiv \delta + 1 =J^{-1},~~~~{\bf u} =
a\left({\partial {\bf x}\over \partial t}\right)_{{\bf q}}= a\dot
{\bf \Psi}. \eqno(7)$$

A solution of Eqs. (6a,b) in the quasi-linear regime may be obtained by
expanding ${\bf \Psi}$ in a power series
${\bf \Psi} = {\bf \Psi}^{(1)} + {\bf \Psi}^{(2)} + ..$ where higher orders
in ${\bf \Psi}^{(n)}$ ($n>1$) are related to lower orders via an iterative
procedure (Moutarde et al. 1991, Buchert 1992, Lachieze-Rey 1993). It
appears that for the matter dominated Universe all ${\bf \Psi}^{(n)}$
factorize:
$$\Psi_i^{(n)}=D_+^n(t)\psi_i^{(n)}({\bf q}),~~~~
D_+(t)= {3t^2\over 2a^2}\propto a(t).\eqno(8)$$
Subsequent Lagrangian approximations then arise if one truncates
this series after a finite number of terms. As a result of this
truncation Eq. (6a) becomes
approximate and the Poisson equation (1a) ceases to be valid, but Eqs.
(7) as well as the continuity equation (1b) continue to be exactly satisfied
in these approximations.

The first-order Lagrangian approximation ${\bf \Psi} ={\bf \Psi}^{(1)}$
is simply the Zeldovich approximation:
$$x_i = q_i + D_+(t)\psi_i^{(1)}({\bf q}),~~~~
\psi_i^{(1)}= -{\partial \phi_0({\bf q})\over \partial q_i}.\eqno(9)$$
The account of the next, second order terms results in what one may call the
{\it post-Zeldovich approximation} (hereafter PZA). It is given by
${\bf \Psi} ={\bf \Psi}^{(1)}+{\bf \Psi}^{(2)}$ where
$$\psi_{i,i}^{(2)} = -{3\over 14}\left((\psi_{i,i}^{(1)})^2
- \psi_{i,j}^{(1)}\psi_{j,i}^{(1)}\right), \eqno(10a)$$
$$\psi_{i,j}^{(2)} = \psi_{j,i}^{(2)}, \eqno(10b)$$
coma means partial derivative with respect to ${\bf q}$ and summation
over repeated indexes is assumed (Bouchet et al. 1992, Gramann 1993
and others). In the third order ({\it post-post-Zeldovich
approximation}, hereafter PPZA),
${\bf \Psi}=\sum_{n=1}^3{\bf \Psi}^{(n)}$ with
$$\psi_{i,i}^{(3)}=  -{5\over 9}(\psi_{i,i}^{(2)} \psi_{j,j}^{(1)}
- \psi_{i,j}^{(2)} \psi_{j,i}^{(1)}) - {1\over 3}det\lbrack \psi_{i,j}^{(1)}
\rbrack, \eqno(11a)$$
$$\psi_{i,j}^{(3)} - \psi_{j,i}^{(3)} = {1\over 3}(\psi_{i,k}^{(2)}
\psi_{k,j}^{(1)} - \psi_{j,k}^{(2)} \psi_{k,i}^{(1)}) \eqno(11b)$$
(cf. Juszkiewicz et al. 1993, Bernardeau 1993) and so on. A distinguishing
feature of PPZA is that motion becomes non-potential in Lagrangian space
(but not in Eulerian space, of course) beginning from this approximation.

Our aim in the present paper will be to determine how well the five nonlinear
approximations discussed above perform when compared within the framework of
perturbation theory which is well defined in the quasi-linear regime when
$|\delta |< 1$. To investigate statistical behaviour in all orders of
perturbation theory we follow Bernardeau (1992) who developed an elegant
ansatz (which we summarise below) by means of which vertice weights
characterising irreducible moments of $\delta$ may be evaluated at every
order by means of a generating function $G_\delta$.

\bigskip
\vfill
\centerline{\bf 3. Generating function for the irreducible moments
of $\delta$.}
\bigskip
\n
Let us introduce a vertice generating function
for any random field $F({\bf x},a)$ as
$$G_F(\tau) = {\sum_{n=1}^{n =\infty}}{\langle F^{(n)}\rangle_c
\over n!} \tau^n \eqno(12a)$$
where
$${\langle F^{(n)}\rangle}_c = {\int \langle F^{(n)}({\bf x}, a)
\delta^{(1)}({\bf x_1}, a)...\delta^{(1)}({\bf x_n}, a)
\rangle_cd^3{\bf x}d^3{\bf x_1}...d^3{\bf x_n}\over
\left(\int \langle \delta^{(1)}({\bf x}, a)\delta^{(1)}({\bf x'}, a)
\rangle d^3{\bf x}d^3{\bf x'}\right)^n},\eqno(12b)$$
$F^{(n)}$ is the n-th order of expansion of $F$ in a power series with
respect to $\delta$, $\delta^{(1)}$ is the linear approximation for
$\delta$ given in Eq. (3), and only connected diagrams are taken into
account, which explains the subscript $c$.
Throughout the present analysis we assume that the initial density
field $\delta^{(1)}$ and the associated linear gravitational potential
$\phi_0({\bf x})$ are Gaussian stochastic quantities though the results
obtained below may be generalized to a non-Gaussian case as well.
Higher vertices of $\delta$ and the dimensionless velocity divergence
$\theta =\triangle U$ are denoted by $\nu_n$ and $\mu_n$ respectively so that
$$
G_\delta = \sum_{n=1}^\infty {\nu_n\over n!} \tau^n,~~~~
G_\theta = \sum_{n=1}^\infty {\mu_n\over n!} \tau^n
\eqno(13)
$$
($\nu_n \equiv \langle\delta^{(n)}\rangle_c, ~~~
\mu_n \equiv \langle\theta^{(n)}\rangle_c$ with $\nu_1=1,\,\mu_1=-1$).
We define $\tau$ with the
opposite sign as compared to Bernardeau (1992) to simplify the
appearance of some expressions.

It is well known that for small values of $\sigma^2 =
\langle \delta^{(1)2}\rangle$ the
connected moments of the density field have the simple form (Fry 1984,
Bernardeau 1992):
$$
\langle\delta^n\rangle_c \simeq S_n\langle\delta^2\rangle^{n-1}
\eqno(14a)
$$
where $S_n$ are related to the vertice weights $\nu_n$ introduced
earlier by
$$S_3 = 3\nu_2$$
$$S_4 = 4\nu_3 + 12\nu_2^2$$
$$S_5 = 5\nu_4 + 60\nu_3\nu_2 + 60\nu_2^3$$
$$S_6 = 6\nu_5 + 120\nu_4\nu_2 + 90\nu_3^2 + 720\nu_3\nu_2^2 + 360\nu_2^4
\eqno(14b)
$$
The equations relating $T_n$ and $\mu_n$ have the same form as (14b)
with $T_n$ replacing $S_n$ and $\nu_n$ being replaced by $(-1)^n\mu_n$.
The additional $(-1)^n$ factor arises because $\mu_n$ are defined with
the use of $\delta^{(1)}$ according to Eq. (12b). Formulas for the moments
of $\theta$ would be completely identical to those for $\delta$ if we
defined $\mu_n$ using $\theta^{(1)}=-\delta^{(1)}$. This just corresponds
to the multiplication of $\mu_n$ by $(-1)^n$.

Now we demonstrate how the vertice weights $\nu_n$ (and hence $S_n$)
can be determined in any arbitrary order for the nonlinear approximations
described above. We do this by following Bernardeau (1992),
who studied leading (tree level) diagrams in the limit when
the density variance $\sigma^2$ was very small and showed that the
generating functions (12a) defined for any arbitrary stochastic fields
$F$ and $H$
possess the following properties:
$$G_{a{ \partial \over \partial a }F}(\tau) =
\tau {\partial \over \partial \tau} G_F,\eqno(15a)$$
$$G_{FH}(\tau)=G_F(\tau)G_H(\tau),\eqno(15b)$$
$$G_{\nabla F\cdot \nabla H} =0,\eqno(15c)$$
$$G_{F_{\alpha\beta}H_{\alpha\beta}}(\tau) = {1\over 3}G_{\triangle F
\triangle G }.\eqno(15d)$$

Using (15a - d) we can rewrite equations (2a,b) in terms of vertice generating
functions of the corresponding fields (as a result, they become
equations for statistically averaged quantities):

$$\tau {d\over d\tau} G_{\delta}  + (1 + G_{\delta})
G_{\theta} = 0,\eqno(16a)$$
$$\tau {d\over d\tau} G_{\theta} + {1\over 2}G _{\theta}
+ {1\over 3}G^2_{\theta} + G_{\triangle \Phi} =0.\eqno(16b)$$
The Poisson equation (2c) is now replaced by:

(1)$G_{\triangle \Phi}={3\over 2}G_{\delta}$ for the exact solution,

(2)$G_{\triangle \Phi}=-{3\over2}G_{\triangle U}= -{3\over2}G_{\theta}$
for ZA,

(3)$G_{\theta}=-\tau$ for FF,

(4)$G_{\triangle \Phi}=G_{\triangle \phi_0}={3\over 2}\tau$ for LP.\\
It is not straightforward to write the corresponding conditions replacing the
Poisson equation for PZA, PPZA and higher Lagrangian approximations
because they are formulated in Lagrangian space, and Eqs. (16a,b) originated
from Eulerian equations. Remarkably, it appears unnecessary because as we
shall show in the next section it is possible to circumvent this problem
entirely.

\bigskip
\centerline{\bf 4. Particle trajectories in the spherical model and
values of the moments $\nu_n, \mu_n$.}
\bigskip
\n
For the exact perturbative solution, the system of equations (16a,b) may be
transformed into a second order differential equation for the variable
$y = 1 + G_{\delta}(\tau)$. For ZA, FF and LP, Eq. (16b) may be solved
directly after substituting the corresponding expressions for $G_{\triangle
\Phi}$. However, a much simpler and more physical method exists that is also
applicable in the cases of PZA, PPZA and higher Lagrangian approximations.

Let us take a closer look at Eqs. (16a,b). It is easy to see that if we make
the substitution $G_{\delta}\to \delta,~G_{\theta}\to \theta$ and $\tau
\to a$, then the statistical Eqs. (16a,b) become dynamical equations
describing the isotropic and homogeneous expansion of
the Universe (a) under the influence of gravity in the case of the
exact solution,
and (b) due to some other forces mimicking gravity for the different
approximations (because the Poisson equation is not satisfied in this case).
The initial conditions $G_{\delta}=\tau,~G_{\theta}=-\tau$ as
$\tau \to 0$ can be satisfied simultaneously if a free scaling dimensional
coefficient of proportionality in dynamical solutions is chosen so that
$\delta = a$ for $a\to 0$. Therefore, the dynamics in question is actually
that of the spherical top-hat model. This agrees well with an intuitive
picture that for $\sigma\ll 1$, large and rare fluctuations
($\sigma\ll \delta <1$) are approximately spherical. The same evidently
refers to the cases of PZA, PPZA and so on, inspite of more complicated
forms for the third equation replacing the Poisson \e, because it follows from
Eqs. (15c,d) that to obtain statistical equations for generating functions
one has to neglect inhomogeneities in dynamical equations and substitute
all second-rank tensors by those proportional to a unit tensor.

Now, the easiest way to solve the top-hat model is to work with Lagrangian
(comoving) coordinates ${\bf q}$. Then the problem reduces to one of finding
appropriate particle trajectories ${\bf x}({\bf q}, a)$ associated with
equations (16a,b) in each of the approximations and in the exact solution
respectively.

\vskip0.4cm

\n
{\it The exact spherical top-hat solution:}
\vskip0.4cm
\n
Let $r=|{\bf x}|,~r_0= |{\bf q}|=r(0)$. The gravitational potential
inside a homogeneous sphere is $ \Phi= {a^2r^2\delta \over 9t^2}$. It clearly
follows from mass conservation that $1+\delta=(r_0/r)^3$. So the equation
of motion of a particle (a spherical shell) is
$${d\over dt}(a^2{dr\over dt})=-{d\Phi \over dr}=-{2a^2r\over 9t^2}
\left(\left({r_0\over r}\right)^3-1\right).\eqno(17)$$
This is none other than the usual Newtonian equation
${d^2R\over dt^2}=-{2a^3\over 9t^2R^2}$ for the reduced physical scale
$R=ar/r_0$. Its first integral satisfying the initial condition $\delta = a$
for $a \to 0$ is
$$\left({dR\over da}\right)^2 = a\left({1\over R}-{5\over 3}\right)
\eqno(18)$$
(the independent variable is changed from $t$ to $a$). The solution of
Eq. (18) in a parametric form is
$$R={3\over 10}(1-cos \theta), \eqno(19a)$$
$$\delta = {9\over 2}{(\theta - sin\theta)^2\over (1 - cos\theta)^3}-1
=G_{\delta},\eqno(19b)$$
$$ a = {3 \over 5}\left({3 \over 4} (\theta - sin \theta)\right)^{2\over 3}
=\tau \eqno(19c)$$
(cf. Zeldovich \& Novikov 1983, Peebles 1980). The corresponding
expression for $G_{\theta}$ follows from Eq. (16a).

The vertice weights $\nu_n$ are found by expanding
$G_\delta(\tau)$ near $\tau = 0$. As a result, we obtain:
$$ G_\delta = \tau + 0.810\tau^2 +0.601\tau^3 + 0.426\tau^4 +
0.293\tau^5 +...,\eqno(20a)$$
$$ G_{\theta}=-(\tau +0.619\tau^2 +0.376\tau^3 +0.226\tau^4 +
0.135\tau^5 +...). \eqno(20b)$$
Comparing (20a) with (13) we find:
$\nu_2 = 34/21 \simeq 1.619,~ \nu_3 = 682/189 \simeq 3.608,~ \nu_4
\simeq 10.23$.
Higher moments of the density distribution such as the skewness $S_3$,
the kurtosis $S_4$ etc. can now be determined by
substituting the values of $\nu_n$ into Eq. (14b).

\vskip0.4cm

\n
{\it The Zeldovich Approximation:}
\vskip0.4cm
\n
In this case, particle trajectories are already given by Eq. (9). The
initial gravitational potential $\phi_0(r_0)$ satisfying the initial
condition $\delta =a$ for $a\to 0$ is $\phi_0=a^3r_0^2/9t^2$ (it doesn't
depend on $t$ since $a(t)\propto t^{2/3}$). As a result
$$r(a) = r_0(1- {a\over 3}),\eqno(21a)$$
$$\delta_{ZA} \equiv G_{\delta} = \left({r_0\over r(\tau)}\right)^3 -1=
\left(1- {\tau \over 3}\right)^{-3} -1, \eqno(21b)$$
$$ G_{\theta} =- \tau \left(1-{\tau \over 3}\right)^{-1}. \eqno(21c)$$
Expanding $G_{\delta}$ and $G_{\theta}$ near $\tau = 0$, we get
$$ G_{\delta} = \tau + 0.667\tau^2 + 0.370\tau^3 + 0.185\tau^4
+0.086\tau^5  +...,\eqno(22a) $$
$$ G_{\theta} =-(\tau +0.333\tau^2 +0.111\tau^3 +0.037\tau^4
+ 0.012 \tau^5 + ...),\eqno(22b)$$
from which we can readily obtain the values of $\nu_n$, $\mu_n$ and $S_n$,
$T_n$. For $n=3,4$, they coincide with those found previously
(Grinstein \& Wise 1987, MS 1993, Bernardeau et al. 1993).

\vskip0.4cm

\n
{\it PZA, PPZA and higher order Lagrangian approximations:}
\vskip0.4cm
\n
Here the Lagrangian method shows its superiority over the Eulerian one
because particle
trajectories are explicitly given by Eqs. (4, 8-10) in the case of PZA,
Eqs. (4, 8-11) for PPZA and so on. Just as in the previous paragraph,
the initial gravitational potential for the top-hat spherical model is
$\phi_0({\bf q})=a^3r_0^2/9t^2$ where $r_0=|{\bf q}|$. Then
$\psi_{i,i}^{(1)}=- 2a^3/3t^2=-a/D_+(t)$,
$D_+^2\psi_r^{(2)}=-a^2r_0/21$ (the index $r$ means a radial component).
Therefore, a particle trajectory in PZA is
$$r(a)= r_0\left(1 - {a\over 3} - {a^2 \over 21}\right),\eqno(23a)$$
and the generating functions for PZA are
$$\delta_{PZA} \equiv G_{\delta} = \left(1 - {\tau \over 3}-
{\tau^2 \over 21} \right)^{-3} -1 = \tau + 0.810\tau^2 +0.561\tau^3
+0.358\tau^4+0.215\tau^5+...,\eqno(23b)$$
$$G_{\theta}=-{\tau +{2\tau^2\over 7}\over 1 - {\tau \over 3}-
{\tau^2 \over 21}}= -(\tau + 0.619\tau^2 +0.254\tau^3
+0.114\tau^4+0.050\tau^5+...).\eqno(23c)$$
In this case, the first two coefficients of the expansion (20a,b) are
reproduced exactly as a result of which PZA gives the correct value for
the skewness $S_3$ in the lowest order in $\sigma$.

Furthermore, $D_+^3\psi_r^{(3)}=-23a^3r_0/1701$ (see Eq. (11a)), so that
a PPZA particle trajectory is given by
$$r(a)= r_0\left(1 - {a\over 3} - {a^2 \over 21}-{23a^3\over 1701}\right).
\eqno(24a)$$
Thus, the generating functions for PPZA have the forms:
$$ G_{\delta} = \left (1 - {\tau \over 3}- {\tau^2 \over 21}
-{23\tau^3\over 1701}\right)^{-3} -1= \tau + 0.810\tau^2 +0.601\tau^3
+0.412\tau^4+0.268\tau^5+..., \eqno(24b)$$
$$G_{\theta}=-{\tau +{2\tau^2\over 7}+{23\tau^3\over189}\over
1 - {\tau \over 3}-{\tau^2 \over 21}-{23\tau^3\over1701}}
=-(\tau + 0.619\tau^2 +0.376\tau^3 +0.168\tau^4+0.082\tau^5+...).\eqno(24c)$$
Now the first three terms in the expansion (20a,b) are correctly
reproduced. Therefore,
the PPZA values for {\it both} the skewness $S_3$ as well as the
kurtosis $S_4$ are exact.

It is clear now that in the $N$-th order Lagrangian approximation defined
by the condition ${\bf \Psi} = \sum_{n=1}^N {\bf \Psi}^{(n)}$, a particle
trajectory in the spherical top-hat model has the form $r(a)=r_0
\left(1-P_N(a)\right)$ where $P_N$ is a $N$-th order polynomial with
$P_N(0)=0$, so the generating functions have the following structure:
$$G_{\delta}=\left(1-P_N(\tau)\right)^{-3}-1, \eqno(25a)$$
$$G_{\theta}=-3{dP_N(\tau) \over d\tau}\left(1-P_N(\tau)\right)^{-1}.
\eqno(25b)$$
The expansion of $G_{\delta}$, $G_{\theta}$ around the point $\tau =0$
correctly reproduces the first $N$ terms
of the series for the exact solution (20a,b) and consequently the first $N+1$
values of $S_n$.

\vskip0.4cm

\n
{\it The Frozen Flow Approximation}
\vskip0.4cm
\n
This is the only case when it is easier to solve the problem in Eulerian
space because the velocity potential is equal to its initial, Gaussian
distributed value. Therefore, $\theta$ is Gaussian, too, and $G_{\theta}
=-\tau$. Consequently
$$G_{\delta}=exp (\tau) -1. \eqno(26)$$

However, the calculation using a particle trajectory method is not long
either. The velocity potential in the top-hat model in FFA is equal to
$U=-2\phi_0(r)/3=-2a^3r^2/27t^2$. So, the equation of motion of a particle
has the form
$$u_r \equiv a{dr\over dt}= -2a^2r/9t.\eqno(27a)$$
Its solution in terms of $a$ is (Matarrese et al. 1992)
$$r(a) = r_0~\exp( - {a\over 3})\eqno(27b)$$
that just gives the expression (26) for the density generating function.

\vskip0.4cm

\n
{\it The Linear Potential Approximation}
\vskip0.4cm
\n
In LP, it is the gravitational potential $\Phi$ that stays equal
to its initial linear value $\phi_0({\bf r})$ which in our case is once
more equal to
$a^3r^2/9t^2$ . The equation of motion of a particle is therefore
$${d\over dt}\left(a^2{dr\over dt}\right)= -{2a^3\over 9t^2}r \eqno(28a)$$
(cf. Eq. (17)) or, it terms of a,
$${d^2r\over da^2}+{3\over 2a}{dr\over da}+{r\over 2a}=0. \eqno(28b)$$
Solving Eq. (28b) we obtain (cf. Brainerd et al. 1993)
$$r ={r_0 \over \sqrt{2a}}\sin\sqrt{2a}.\eqno(29a)$$
Therefore,
$$G_{\delta} = \left({ \sqrt{2\tau }\over
sin\sqrt{2 \tau}}\right)^3 -1= \tau + 0.567\tau^2 +0.242\tau^3 +
0.087\tau^4 +0.028\tau^5 +...,\eqno(29b)$$
$$G_{\theta} ={3\over 2}\left(\sqrt{2\tau}cot\sqrt{2\tau}-1\right)=
-(\tau +0.133\tau^2 +0.025\tau^3 +0.0051\tau^4 +0.0010\tau^5 +...)
\eqno(29c)$$
in LP.

\vskip0.4cm

In figure 1 we plot the density contrast in the spherical top-hat model
obtained using PPZA, PZA, ZA, LP and FF against the exact solution.
We find that for a positive density perturbation all models underestimate
the density contrast, but PPZA, PZA and ZA are more accurate than LP and FF.
This is quite surprising since it is well known that the Zeldovich
approximations is the least accurate during spherical collapse when all
eigenvalues of the deformation tensor
$\partial^2U/\partial q_i\partial q_j$ are equal
(Shandarin $\&$ Zeldovich 1989). We also list the turnaround epoch and the
recollapse epoch (in units of $\tau$) in each of the approximations and in the
exact solution in Table 1. We find that both turnaround and recollapse occur
later in the approximations than in the exact solution.

Expanding $G_{\delta}$ and $G_{\theta}$
in each of the approximations near $\tau = 0$ we
get the reduced moments $\nu_n$ $\mu_n$ and, from (14b), the parameters $S_n$
and $T_n$.
Our results are summarised in Table 2 for the first six moments of the
distribution: $S_1 ... S_6$. Table 3 contains the values of $T_1 ... T_6$.
The values which we obtain for $S_3$ and $S_4$
for ZA, FF and LP are identical to those
obtained earlier by MS and Bernardeau et al. (1993).
Interestingly we find that of all the approximations considered by us,
PPZA appears to be the most accurate in reproducing the results of the exact
perturbative treatment. Next in accuracy comes PZA which reproduces the
skewness only, then ZA followed by LP and FF.
The results of our analysis lead us to conclude that nonlinear approximations
formulated in Lagrangian space (ZA, PZA etc.) are significantly more accurate
in the weakly nonlinear regime than those which are formulated in
Eulerian space (FF and LP).
It would be of great interest to extend this analysis to the strongly
nonlinear regime where the different approximations should be compared
with results of the adhesion model and N-body simulations.
Some work in this direction is presently in progress (Sathyaprakash
et al. 1994).

\bigskip
\centerline{\bf 5. Density distribution functions}
\bigskip

\n
Having obtained the generating functions $G_{\delta}$, it is fairly
straightforward to calculate corresponding density PDFs $\eta(\delta)$
for each of the approximations in the limit $\sigma \ll 1$ and for a
sufficiently small
$|\delta|$ (for the exact perturbative solution this was done by
Bernardeau (1992)). A more detailed discussion of the region of
applicability of the approach used in the paper will be given below.

Let ${\cal P}(y)$ be the Laplace transform of $\eta(\delta)$ (with $\sigma$
displayed explicitly):
$${\cal P}(y)=\int_{-1}^{\infty}\eta(\delta)exp\left(-{(1+\delta)y
\over \sigma^2}\right)\, d\delta\,.\eqno(30)$$
Then the generating function of the moments $S_n$ (see Eq. (14a)):
$$\varphi (y) = \sum_{p=1}^{\infty}S_p{(-1)^{p-1}\over p!}y^p,~~S_1=S_2=1,
\eqno(31)$$
is connected to ${\cal P}$ by the simple formula ${\cal P}(y)=
exp(-\varphi(y)/\sigma^2)$ (see, e.g., White 1979). On the other hand, as
can be checked by a direct comparison of the coefficients of the series (13)
and
(31) using (14b) that $\phi (y)$ can be expressed through the function
$\zeta(\tau)\equiv 1+G_{\delta}(\tau)$ using the following relations:
$$ \varphi (y)=y\zeta\left(\tau (y)\right)+{1\over 2}\tau^2,\eqno(32a)$$
$$\tau(y)=-y\dot \zeta\left(\tau(y)\right) \eqno(32b)$$
where dot means derivative with respect to $\tau$ (see, e.g., Bernardeau
\& Schaeffer 1992), so that
$${d\varphi(y)\over dy}=\zeta\left(\tau(y)\right),~~~{d^2\varphi(y)\over
dy^2} =\dot \zeta {d\tau\over dy}=-{\dot\zeta}^2\left(1-{\tau \ddot \zeta
\over \dot \zeta}\right)^{-1}.\eqno(32c)$$
Finally, the density PDF follows from the inverse Laplace transformation
$$\eta(\delta)={1\over 2\pi i\sigma^2}\int_{c-i\infty}^{c+i\infty}
exp\left({(1+\delta)y-\varphi(y)\over \sigma^2}\right)dy \eqno(33)$$
where $c=const$.

For example, for a purely Gaussian stochastic field, $\zeta(\tau)=
1+\tau,~y=-\tau$ and $\varphi (y)=y-y^2/2$ in accordance with the fact that
all $S_p$ with $p > 2$ in Eq. (31) are equal to zero in this case. Then the
inverse Laplace transform of this $\varphi(y)$ is just the Gaussian
distribution
$\eta(\delta)=(2\pi\sigma^2)^{-1/2}exp\left(-\delta^2/2\sigma^2\right)$.

In the limit $\sigma \ll 1$, the main contribution to the integral in (33)
is made by stationary points of the exponent in it, i.e. by roots of the
equation
$${d\varphi (y)\over dy} \equiv 1+G_{\delta}(\tau)=1+\delta \eqno(34)$$
(complex roots of this equation should be considered, too). Thus, here
we return to the prescription $\delta \to G_{\delta}$ used in the derivation
of equations for the generating functions in Sec. 3. In this (steepest
descent) approximation, Eq. (33) takes the form
$$\eta (\delta)={1\over \sqrt{2\pi \sigma^2}}{1\over \dot \zeta}
\sqrt{1-{\tau \ddot \zeta \over \dot \zeta}} exp\left(-{\tau^2\over
2\sigma^2}\right) \eqno(35)$$
where now $\tau(\delta)$ should be determined from Eq. (34) for any given
generating function $G_{\delta}(\tau)$.

Let us now discuss the accuracy of the formulae (33, 35) for the
density PDF. Since the generating
functions $G_{\delta}(\tau)$ and $\varphi(y)$ have been calculated in the
lowest order in $\sigma$ only, there exist small corrections to them
beginning from terms proportional to $\sigma^2$ so that $\varphi(y)=
\varphi_0(y)+\sigma^2\varphi_1(y)+...$ (here the zero-order term
$\varphi_0(y)$ stands for $\varphi(y)$ used above) and similarly for
$G_{\delta}$. The account of the $\sigma^2$ correction results in the
multiplication of the integrand of Eq. (33) by the
$\sigma$-independent term $exp(-\varphi_1(y))$ having an unknown dependence
on $y$. Similarly, the multiplicative factor
$exp(-\varphi_1(y(\tau)))$ will appear
in the right-hand side of Eq. (35). In the limit $|\delta|\ll 1$ when
$\delta \approx\tau \approx -y$, this factor has the form
$(1+O(\delta^2))$, the $\delta^2$ (or $\tau^2$) correction arising due
to a $\sigma^4$ correction to the dispersion of density perturbations.
Of course, we may renormalize $\sigma^2$ by defining it to be equal to
the {\it exact} dispersion of density perturbations and not
$\langle\delta^{(1)2}\rangle$ as is used in the paper, but all other
corrections beginning from a term $\propto \delta^3$ will remain.

Therefore, we have to conclude that if each term in the power series (13)
for the generating function $G_{\delta}$ is computed in the lowest order in
$\sigma$ as is done in Bernardeau 1992 and in the present paper, then the
resulting density PDF has an exponential accuracy: the large exponential factor
in it appears to be correct for not too large $\delta$, but the coefficient
of the exponential should be taken in the limit $\delta \to 0$.
One is allowed to keep terms linear in $\delta$ in the latter but we shall not
do so except for retaining the coefficient $1/\sqrt{2\pi \sigma^2}$ in front
of the large exponential (we make an exception for the Zeldovich approximation
where it is instructive to keep the entire coefficient in Eq. (35) in
order to
compare it with an exact analytic solution). This is enough, however, to obtain
very significant deviations from Gaussian behaviour in the quasi-linear
regime.

On the other hand, to go beyond exponential accuracy and evaluate the
coefficient of the exponential exactly, one has to calculate $\varphi_1(y)$.
Generally, this quantity is spectrum-dependent although the ZA presents
an important exception to this rule.
But even then, the resulting expression may not be used for arbitrarily
large $\delta$ in the limit $\sigma \ll 1$. In particular, it is
not possible to get the threshold value $\delta_c$ appearing in a
Press-Schechter-like formula for the number of collapsed objects from such
a treatment. This
arises because other density configurations different from the spherical
top-hat one may give a dominant contribution to the generating function
$G_{\delta}$ in this regime. It results for instance,
in that the value $\delta_c$
for ZA should be taken as $\sqrt5$ in the limit $\sigma \ll 1$ as shown
in the Appendix, and not as $3$ as seems from Eq. (38) below. Moreover, the
case of the Zeldovich approximation shows that the  value of $\delta$ for
which such a new dominant configuration appears cannot be obtained from
the spherical approximation for the vertice generating function.

Now we consider the concrete approximations.

\bigskip

\n
{\it The Zeldovich approximation:}\\
\n
This case is very important because we may compare our result with an
exact analytic solution for the density PDF valid for all values of $\sigma$
and $\delta$ which was obtained by Kofman (1991). This permits us
to check the range of validity of the method used. This solution and
its limit for $\sigma \ll 1$ are given in the Appendix.

Using Eqs. (21b, 32a,b, 33) and changing the integration variable in (33)
from $y$ to $\tau$, we obtain
$$\zeta(\tau)=\left(1-{\tau \over 3}\right)^{-3},~~~y=-\tau \left(1-
{\tau \over 3}\right)^4,~~~\varphi(y)=-\tau+{5\over 6}\tau^2, \eqno(36a)$$
$$\eta(\delta)=-{1\over 2\pi i\sigma^2}\int_C\left(1-{\tau \over 3}\right)^3
\left(1-{5\tau \over 3}\right)exp\left({\tau-{5\over 6}\tau^2-(1+\delta)
\tau\left(1-{\tau \over 3}\right)^4 \over \sigma^2}\right)d\tau, \eqno(36b)$$
where the integration contour C corresponding to that in Eq. (33) is a
continuous curve in the complex $\tau$ plane beginning in the sector $|\tau|
\to \infty,~~0.7\pi < arg(\tau) < 0.9\pi$ and ending in the sector $|\tau|\to
\infty, ~~-0.9\pi < arg(\tau) < -0.7\pi$. The only irregular non-analytic
point of the integrand is at $|\tau|=\infty$. So, for $\sigma \ll 1$, only
stationary points of the exponent should be considered. There are four of them:
$$\tau_n=3\left(1-{e^{i\beta_n}\over (1+\delta)^{1/3}}\right),~~~n=1,2,3,~~~
\beta_n={2\pi (n-1)\over 3},~~~\tau_4=0.6.\eqno(37)$$
The last point does not make a contribution in the lowest order in $\sigma$
because of the coefficient in front of the exponential in Eq. (36b). The point
$\tau_1$ gives the largest exponent for not too large $\delta$. The direction
of the steepest descent for it is perpendicular to the real axis. Thus,
we find using (35) that
$$\eta(\delta) = {1\over \sqrt{2\pi\sigma^2}}\left(1-{\tau \over 3}
\right)^{7/2}\left(1-{5\tau \over 3}\right)^{1/2}e^{-{\tau^2\over 2\sigma^2}},
{}~\tau(\delta)=3\left(1-{1\over (1+\delta)^{1/3}}\right).\eqno(38)$$

Comparing this expression with the rigorous result for ZA in the limit
$\sigma \ll 1$ (Eq. (A.2)), we see that the large exponential is reproduced
exactly up to a rather large value of $\delta$ but the coefficient of the
exponential in Eq. (38) is correct in the limit of small $|\delta|$ only.
In the latter limit, Eq. (38) just reduces to Eq. (A.3) if terms up to
$O(\delta)$ inclusive are kept in the coefficient. $O(\delta^2)$ and
higher terms in series expansions of the coefficients of the
exponential in (38) and (A.2) are different. Moreover, the appearance of the
square root singularity in Eq. (38) at $\tau =0.6,~~\delta \approx 0.95$
is an artifact of the approach used, this point is completely regular in the
rigorous density PDF for ZA. All this is in complete agreement with the
general discussion of limits of validity of the spherical approximation
for the vertice generation function in the beginning of this section.
(Note also that the exponent in Eq. (38) coincides with that found by
Padmanabhan \& Subramanian (1993) for the PDF of the {\it final
smoothed} density field in ZA using a completely different approach.)

\bigskip

\n
{\it PZA and PPZA:}\\
\n
With the accuracy chosen above, we immediately get from Eqs. (23b,
24b, 34, 35) that
$$\eta(\delta)\approx  {1\over \sqrt{2\pi \sigma^2}}
exp\left(-{\tau^2(\delta)\over 2\sigma^2}\right)\eqno(39)$$
where
$$\delta(\tau) = \left (1 - {\tau \over 3}- {\tau^2 \over 21}
\right)^{-3} -1 \eqno(39a)$$
for PZA and
$$ \delta(\tau) = \left (1 - {\tau \over 3}- {\tau^2 \over 21}
-{23\tau^3\over 1701}\right)^{-3} -1 \eqno(39b)$$
for PPZA.

\bigskip

\n
{\it The Frozen flow approximation:}\\
\n
Now it follows from Eqs. (26, 34, 35) that $\tau =ln(1+\delta)$ and
$$\eta(\delta)\approx {1\over \sqrt{2\pi\sigma^2}}{1\over 1+\delta}exp\left
(-{ln^2(1+\delta)\over 2\sigma^2}\right),\eqno(40)$$
where we  have introduced the coeficient $(1+\delta)^{-1}$ (that is possible
within our accuracy) to make the PDF being normalized to $1$ exactly.
PDF (40) is just the log-normal distribution which was
independently proposed as a good {\it statistical} approximation for the
density PDF in the quasi-linear regime by Hamilton (1988) and Coles \&
Jones (1991) (in contrast to the {\it dynamical} approximations which we
are considering).
The result (40) shows that there exists a close internal relationship between
FFA and the log-normal approximation. In particular, we may expect that
they have approximately the same accuracy in the quasi-linear regime.

\bigskip

\n
{\it The Linear Potential approximation:}\\
Eqs. (29b, 34, 35) are now relevant, giving the result
$$\eta(\delta)\approx  {1\over \sqrt{2\pi \sigma^2}}
exp\left(-{\tau^2(\delta)\over 2\sigma^2}\right),~~~
\delta(\tau) = \left({ \sqrt{2\tau }\over
sin\sqrt{2 \tau}}\right)^3 -1. \eqno(41)$$

\bigskip

\centerline {\bf 6. Conclusions}
\vskip 0.4cm

\n
The nonlinear evolution of an initially Gaussian distribution leads to the
following relationship between the connected density moments
$\langle\delta^{(n)}\rangle_c$ at the lowest order in $
\langle\delta^2\rangle$:
$ \langle\delta^n\rangle_c \simeq S_n\langle\delta^2\rangle^{n-1}$
where the parameters $S_p$ characterise the development of non-Gaussianity
($S_3$ describes the skewness of the density distribution and $S_4$
its kurtosis).
Generalising earlier work by Bernardeau (1992) we have shown that
for approximation methods attempting to mimick the effects of nonlinear
gravity
the values of $S_p$ can be derived by means of
a generating function $G_{\delta}$. For a given nonlinear approximation
$G_{\delta} \equiv \delta_{sph}$ is simply the overdensity in the spherical
top hat model in that approximation. Thus knowing how particle trajectories
evolve in the spherical top hat model in a given nonlinear approximation
we can determine $G_{\delta}$ and consequently $S_p$ and the probability
distribution function for the density.

Following this ansatz we determine the functional form of
the generating function and the values of the
first six parameters $S_1, ... S_6$ for five distinct nonlinear approximations,
of which three are formulated in Lagrangian space (including the Zeldovich
approximations and its extensions), and the remaining two in Eulerian space
(frozen flow and linear potential). Comparing our results with those of an
exact perturbative treatment (Bernardeau 1992) we find that an approximation
formulated to $n^{th}$ order in Lagrangian space
correctly reproduces the first
$n + 1$ parameters $S_n$ (see Table 2).
Our comparison leads us to conclude that nonlinear
approximations which are formulated in Lagrangian space are considerably more
accurate than those formulated in Eulerian space when tested in the weakly
nonlinear regime.

\vskip 0.4cm
\leftline{{\bf Acknowledgements}}

\n
The authors are grateful to F. Bernardeau, L. Kofman, T. Padmanabhan
and B.S. Sathyaprakash
for stimulating discussions.  A.A.S. is grateful to Profs. Y. Nagaoka and
J. Yokoyama for their hospitality at the Yukawa Institute for
Theoretical Physics. The financial support for the research work of
A.A.S. in Russia was provided by the Russian Foundation for Basic Research,
Project Code 93-02-3631. D.M. was financially supported by the Council of
Scientific and Industrial Research, India, under its JRF scheme.

\eject

\bigskip
\bigskip
\centerline{\bf APPENDIX}
\bigskip
\n
The rigorous formula for the density PDF in the Zeldovich approximation for
a single-stream motion in Eulerian space with Gaussian initial
conditions is (Kofman 1991, see also Kofman et al. 1994)
$$\eta(\delta)={9\cdot 5^{3/2}\over 4\pi \sigma^4(1+\delta)^3}
\int_{s_0(\delta)}^{\infty} e^{-{(s-3)^2\over 2\sigma^2}}
\left(1+e^{-{6s\over \sigma^2}}\right)\left(e^{-{\beta_1^2\over 2\sigma^2}}
+e^{-{\beta_2^2\over 2\sigma^2}}-e^{-{\beta_3^2\over 2\sigma^2}}\right)ds,
\eqno(A.1a)$$
$$s_0(\delta)={3\over (1+\delta)^{1/3}},~~\beta_n(s,\delta)=s\cdot
\sqrt 5 \left({1\over2}+cos\left( {2\over 3}(n-1)\pi +{1\over3}arccos
\left(2\left({s_0\over s}\right)^3-1\right)\right)\right). \eqno(A.1b)$$

For $\sigma \ll 1$ and not too large $\delta$, the main contribution to the
integral comes from the lower limit of integration $s=s_0$. If $s=s_0(1+x),~~
x\ll 1$, then $\beta_1=s_0\cdot 3\sqrt 5/2 +O(x),~~\beta_{2,3}=\mp s_0
\cdot \sqrt{5x}\left(1\mp \sqrt x/3+O(x)\right)$. Then the terms with
$\beta_{2,3}$ are the leading ones (they almost cancel each other),
and the formula (A.1a) takes the form
$$\eta(\delta)={1\over \sqrt{2\pi \sigma^2}}{\left(1-{\tau\over3}\right)
^{19/2}\over \left(1-{7\tau\over 15}\right)^{5/2}}e^{-{\tau^2 \over
2\sigma^2}},~~~\tau =3\left(1-{1\over (1+\delta)^{1/3}}\right). \eqno(A.2)$$
If $|\delta| \ll 1$, too (but $|\delta|$ may be much more than $\sigma$),
then (A.2) simplifies to
$$\eta(\delta)= {1-2\delta+O(\delta^2)\over \sqrt{2\pi \sigma^2}}
exp\left(-{9\over 2\sigma^2}\left(1-{1\over (1+\delta)^{1/3}}\right)^2
\right). \eqno(A.3)$$

The expression (A.2) clearly looses sense for $\delta \ge (7/2)^3-1
\approx 42$, but it is no more the dominant term in (A.1a) for this value and
even for somewhat smaller values of $\delta$. A numerical calculation shows
that for $\delta \ge 30.6$,  the main contribution to the PDF is produced
by the maximum of the exponential $exp\left(-\left((s-3)^2+\beta_2^2
\right)/2\sigma^2\right)$ which is located at $s\approx 4/3$ for $\delta\to
\infty$, other terms in (A.1a) being exponentially smaller. As a result,
$\eta(\delta)\propto (1+\delta)^{-3}exp(-\delta_c^2/2\sigma^2)$ with
$\delta_c=\sqrt 5$ for $\delta \to \infty$ and $\sigma \ll 1$. Thus, the
Press-Schechter-like density perturbation value $\delta_c$ (a ``threshold''
for formation of compact objects)
is equal to $\sqrt 5$ in the Zeldovich approximation, and not
to $3$ that would follow from a naive application of the formula (A.2)
in the regime $\delta \gg 1$. Of course, even $\sqrt 5$ is a great
overestimation of a real value of $\delta_c$ in the exact solution (if
it has sense at all) which is expected to be equal to, or a little bit less
than $1.686$.

Note also that the PDF (A.1a) is not normalized to unit total probability
due to the appearance of multistreaming motion in the Zeldovich
approximation (as well as in the exact solution) even for a small $\sigma$.
Actually, $\int_{-1}^{\infty}\eta(\delta) d\delta$ gives the mean number
of streams $N_s$ (see also Kofman et al. 1994). This is, however, an
exponentially small effect, it is straightforward to show by direct
integration of Eq. (A.1a) that
$$N_s =1+{16\over 27\sqrt{10\pi}}~\sigma \exp \left(-{5\over 2\sigma^2}
\right) \eqno(A.4)$$
for $\sigma \ll 1$. Here, the threshold value $\delta_c =\sqrt5$ appears
once more. This formula is not bad even for $\sigma =1$ where it gives
$N_s\approx 1.0087$ instead of the exact value $N_s(1)\approx 1.0137$.

\vfill

\begin{table*}
\begin{center}
\begin{tabular}{lcccccc} \hline \hline
  & ~~~Exact~~~  &  ~~~PPZA~~~  &  ~~~PZA~~~ &  ~~~ZA~~~ & ~~~LP~~~ &  ~~~FF~~~
 \\ \hline

$\tau_{ta}$  &  1.062  &     1.118 &    1.194 &   1.5 &  2.058 &  3  \\

$\tau_{coll}$   &  1.686   &   2.050   &  2.266   &  3.      &  4.935 &
$\infty$  \\
\hline

\end{tabular}
\end{center}
\caption{ Spherical collapse}
\label{tbl-1}
\end{table*}

\begin{table*}
\begin{center}
\begin{tabular}{lcccccc} \hline \hline
 & ~~~Exact~~~  &  ~~~PPZA~~~  &  ~~~PZA~~~ &  ~~~ZA~~~ & ~~~LP~~~ &  ~~~FF~~~
\\ \hline

$S_3$  &  4.857  &      4.857 &       4.857 &   4. &      3.4 &  3  \\

$S_4$   &  45.89   &    45.89   &  44.92   &  30.22      &  21.22 & 16  \\

$S_5$   &  656.3  &     654.6  &    624.4  &  342.2   &    196.4 & 125  \\

$S_6$  &  12,653       &  12,568   &  11,666   &  5200     &  2429 & 1296  \\
\hline

\end{tabular}
\end{center}
\caption{ Moments of $\delta$ field}
\label{tbl-2}
\end{table*}

\begin{table*}
\begin{center}
\begin{tabular}{lcccccc} \hline \hline
& ~~~Exact~~~  &  ~~~PPZA~~~  &  ~~~PZA~~~ &  ~~~ZA~~~ & ~~~LP~~~ &  ~~~FF~~~
\\ \hline

$T_3$  &  -3.714  &   -3.714 &   -3.714 &  -2  &     -0.8  &  0.  \\

$T_4$   &  27.41   &    27.41   &  24.49   &  8.      &  1.46 & 0.  \\

$T_5$   &  -308.4  &    -301.5  &   -240.8  &  -48.9   &   -4.19 & 0.  \\

$T_6$  &  4694       &  4450   &  3180   &  404.4     &  16.35 & 0.  \\ \hline

\end{tabular}
\end{center}
\caption{ Moments of $\theta$ field}
\label{tbl-2}
\end{table*}

\eject

\bigskip

\bigskip
\eject
{\bf Figure Captions}

\medskip

{\bf Fig. 1} : The density in the top-hat spherical collapse model as
estimated in the different approximations ($y(\tau) = 1 + \delta$)
is plotted against the exact solution.
The exact solution is labelled by 1; PPZA by 2; PZA by 3; ZA by 4; LP by
5;
FF by 6; and the linear solution by 7.
\medskip

%
\medskip


\end{document}